\documentclass{PoS}

\title{The rise and fall of a binary AGN candidate: the story of PSO J334.2028+1.4075}

\ShortTitle{Binary AGN candidate PSO J334.2028+1.4075}

\author{P. Benke,$^a$ \speaker{S. Frey},$^{b}$ K. \'E. Gab\'anyi,$^{cb}$ L. I. Gurvits,$^{de}$ Z. Paragi,$^{d}$ T. An,$^{f}$ E. Kun,$^{g}$ P.~Mohan,$^{f}$ D. Cseh$^{b}$ and Gy. Mez\H{o}$^{b}$\\
\llap{$^a$}Department of Astronomy, E\"otv\"os Lor\'and University, Budapest, Hungary\\
\llap{$^b$}Konkoly Observatory, MTA Research Centre for Astronomy and Earth Sciences, Budapest, Hungary\\
\llap{$^c$}MTA-ELTE Extragalactic Astrophysics Research Group, Budapest, Hungary\\
\llap{$^d$}Joint Institute for VLBI ERIC, Dwingeloo, The Netherlands\\
\llap{$^e$}Department of Astrodynamics and Space Missions, Delft University of Technology, The Netherlands\\
\llap{$^f$}Shanghai Astronomical Observatory, Chinese Academy of Sciences, Shanghai, China\\
\llap{$^g$}Institute of Physics, University of Szeged, Hungary\\
E-mail: \email{benkepetra@caesar.elte.hu}, \email{frey.sandor@csfk.mta.hu}, \email{krisztina.g@gmail.com}}

\abstract{Apparently periodic optical variations of the luminous high-redshift ($z=2.06$) quasar PSO~J334.2028+1.4075 led Liu et al. (2015) to interpret the variability as the orbital period of a binary supermassive black hole (SMBH) residing in a single circumbinary accretion disk. The proposed orbital separation was around 0.006~pc, and the possible inspiral time about 7 yr in the rest frame of the quasar. Such objects would be of high interest as the difficult-to-find end products of binary SMBH evolution, and potential sources of low-frequency gravitational waves. However, extending the time baseline of the variability study, Liu et al. (2016) later found that the periodicity of PSO~J334.2028+1.4075 does not remain persistent. Foord et al. (2017) did not find evidence for the binary active galactic nucleus scenario based on \emph{Chandra} X-ray observations. The object has also been studied in detail in the radio (Mooley et al. 2018) with the Karl G. Jansky Very Large Array (VLA) and the Very Long Baseline Array (VLBA), revealing a lobe-dominated quasar at kpc scales, and possibly a precessing jet, which might retain PSO~J334.2028+1.4075 as a binary SMBH candidate. Here we report on our 1.7-GHz observation with the European VLBI Network (EVN) which complements the VLBA data taken at higher frequencies, and discuss the current knowledge about the nature of this interesting object.}

\FullConference{14th European VLBI Network Symposium \& Users Meeting (EVN 2018)\\
		8-11 October 2018\\
		Granada, Spain}

\begin{document}

\section{Introduction}

Since it is extremely rare to find supermassive black hole binaries (SMBHBs), the discovery of PSO~J334.2028+1.4075 (FBQS~J2216+0124; PSO~J334 hereafter) \cite{Liu15} attracted considerable interest. Supermassive black holes usually reside in large elliptical galaxies. Their mass could reach $\sim10^{10}$~M$_\odot$. Although SMBHBs might seem as common objects due to galaxy mergers, it is problematic to detect them, especially at small separations \cite{An18}. The SMBHB candidate quasar PSO~J334 was selected via a systematic search in the data of the Pan-STARSS1 Medium Deep Survey \cite{Liu15}. Based on the observed period in the variation of the optical flux ($\sim 542$ days) and the estimated total black hole mass ($\sim 10^{10}$~M$_\odot$, with a mass ratio of $0.05<q<0.25$), an orbital separation of 0.006~pc was inferred. According to this, the coalescence of the SMBHB would occur in approximately 7 yr in the rest frame of the quasar \cite{Liu15}. Unfortunately, none of our instruments are capable of resolving the two components at such a small separation. So the evidences of the second component could only be indirect, like the detected variation in the optical flux. This could be caused by a secondary black hole passing through the primary black hole's accretion disk, as proposed for OJ~287 \cite{Lehto96}.

However, the detected 2.6 cycles of the putative periodicity is likely insufficient to claim sinusoidal variations \cite{Vaughan16}. Indeed, as a result of a more detailed examination involving extended time baselines of optical monitoring, no proof of periodic variability was found in PSO~J334 \cite{Liu16}.

PSO~J334 is a radio source and its structure has been investigated at several radio frequencies with the Karl G. Jansky Very Large Array (VLA) and the Very Long Baseline Array (VLBA) interferometers \cite{Mooley18}. According to the VLBA images obtained at four frequencies from 4.4 to 15.4~GHz, the quasar is resolved into two components, a compact core and a jet. Their distance is 3.6 milliarcseconds (mas), corresponding to 30~pc projected linear separation \cite{Mooley18}. According to the VLA images, the extended jet and lobe structure is stretching 66~kpc from the opposite sides of the core. What is more, the $39^{\circ}$ difference between the position angles of the outer lobes and the inner jet is so significant that it could suggest a second SMBH's disturbing effect on the jet. So, despite the outcome of the most recent optical light curve analysis, PSO~J334 can still be considered a SMBHB candidate \cite{Mooley18}. 
Multi-waveband observations aiming to determine the accretion mode of the quasar \cite{Foord17} did not find any feature that would convincingly distinguish PSO~J334 from a single active galactic nuclus (AGN). However, there are still scenarios allowing that the object is a SMBHB.

We studied the radio structure of PSO~J334 with the technique of very long baseline interferometry (VLBI) using the European VLBI Network (EVN) at 1.7~GHz. Here we present our EVN image, the measurement of the 1.7-GHz flux density of the source, and compare our results with those obtained with the VLBA at higher frequencies \cite{Mooley18}.

\section{EVN observation, data analysis and results}

Eleven radio telescopes provided data in our EVN experiment (project code RSG08, PI: S.~Frey) conducted on 2015 October 18: Jodrell Bank Lovell Telescope (United Kingdom), Westerbork (single dish; the Netherlands), Effelsberg (Germany), Medicina (Italy), Onsala (Sweden), Sheshan (China), Toru\'{n} (Poland), Hartebeesthoek (South Africa), Svetloe, Zelenchukskaya, and Badary (Russia). The correlation was performed at the Joint Institute for VLBI ERIC (Dwingeloo, the Netherlands) with 4~s integration time. The observations lasted for 2~h. The data were recorded at 1024~Mbit~s$^{-1}$ rate in both left and right circular polarizations, with 8 basebands per polarization, each divided into thirty-two 500-kHz wide spectral channels. 
During the observation, we used phase-referencing to a nearby (within $1^{\circ}$) compact calibrator source J2217+0220, with duty cycles of 6.5~min. 
The total accumulated observing time on PSO~J334 was 1~h. The data were calibrated with the NRAO Astronomical Image Processing System (AIPS) \cite{Greisen03} following the usual procedures. Imaging and brightness distribution model fitting was done in Difmap \cite{Shepherd97} in the standard way. 

\begin{figure}
\center
	\includegraphics[width=0.40\columnwidth]{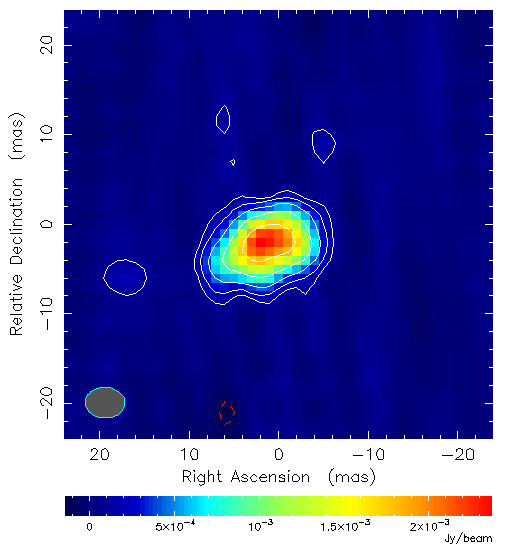}
        \includegraphics[width=0.59\columnwidth]{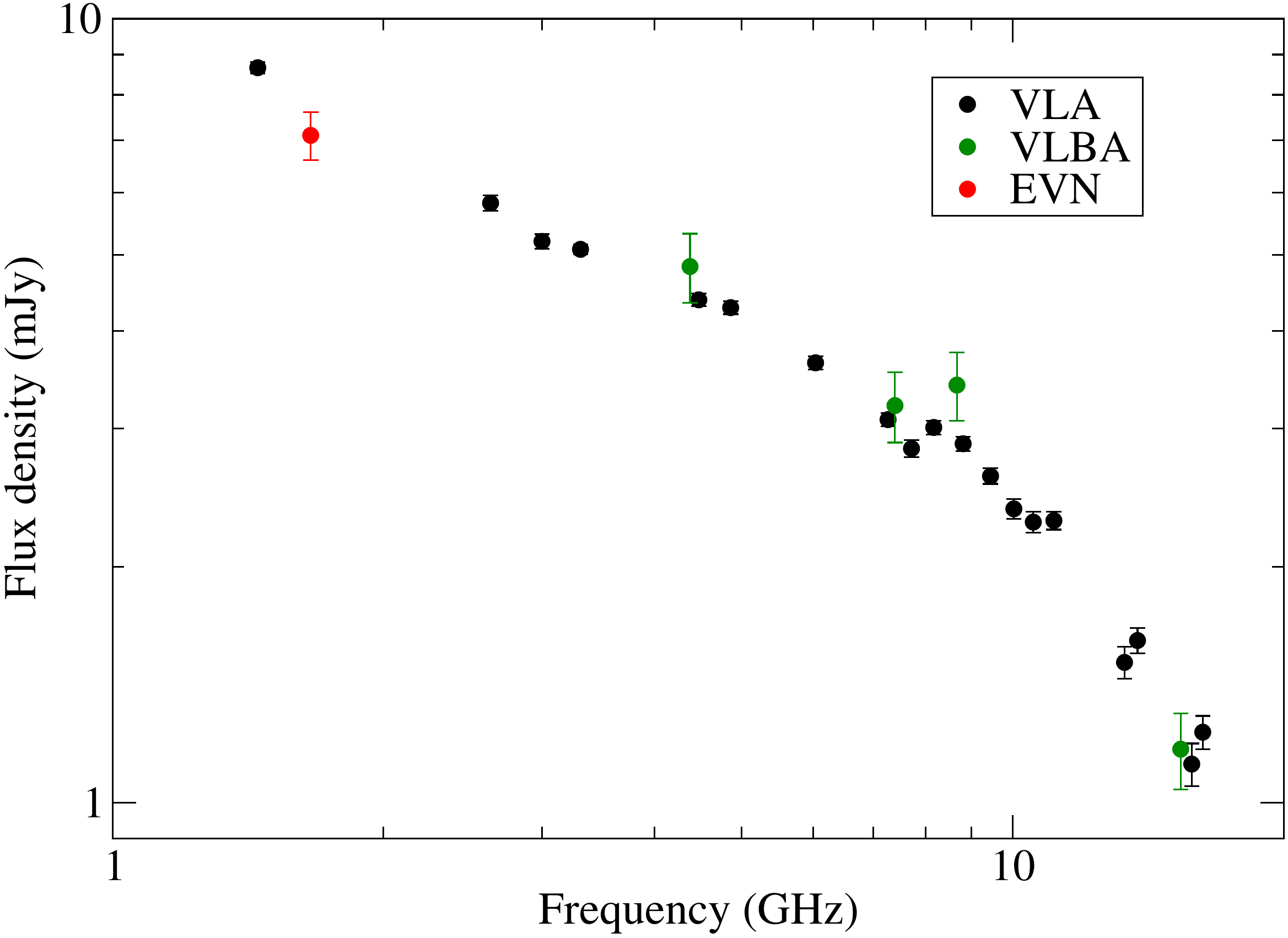}
	\caption{\emph{Left:} 1.7-GHz EVN image of PSO~J334 (J2216+0124). The peak brightness is 2.38~mJy~beam$^{-1}$. The lowest contours are drawn at $\pm 0.109$~mJy~beam$^{-1}$ corresponding to $3\sigma$ image noise level. The positive contour levels increase by a factor of 2. The Gaussian restoring beam shown in the lower-left corner is 4.4~mas $\times$ 3.4~mas (full width at half maximum) with a major axis position angle $89^\circ$. \emph{Right:} Radio spectrum of PSO~J334 based on non-simultaneous VLA and VLBA measurements \cite{Mooley18}. The 1.7-GHz flux density measurement from our EVN experiment is shown in red.}
	\label{evn}
\end{figure}

The 1.7-GHz EVN image of PSO~J334 (Fig.~\ref{evn}, left) with an angular resolution of $\sim 4$~mas shows a single component that is slightly resolved, roughly in the east--west direction. The structure is consistent with the higher-frequency VLBA images \cite{Mooley18} where two components -- a south-eastern synchrotron self-absorbed core and a north-western jet -- were identified with 3.6~mas separation along the position angle $139^{\circ}$ (measured from north through east).

We fitted a simple circular Gaussian brightness distribution model component to the interferometric visibility data, to obtain an estimate of the flux density in the compact radio structure at 1.7~GHz. The value $S=7.1 \pm 0.5$~mJy is added to the radio spectrum compiled in \cite{Mooley18} and is displayed in the right panel of Fig.~\ref{evn}. The new measurement fits well in the overall power-law spectrum. Our value is somewhat below what would be expected from the interpolation between the neighbouring spectral points. However, those are VLA measurements with lower anguluar resolution, and flux density variability is also possible. 

We also observed PSO~J334 with the VLA in its A configuration on 2016 October 26. The detailed analysis and results will be presented elsewhere (Gab\'anyi et al., in prep.). The new 6-GHz VLA image reveals a remarkably straight jet up to $\sim 5^{\prime\prime}$ on the eastern side of the core where it bends sharply and terminates in a hot spot. This structure seems hard to reconcile with the jet precession \cite{Mooley18}. It rather suggests an arcsec-scale jet possibly interacting with and deflected by the dense ambient medium (e.g., as in the radio galaxy 4C~41.17 \cite{Gurvits97}). There is no evidence for a helical structure of a precessing jet. This, and recent studies from the literature \cite{Vaughan16,Liu16,Foord17} indicate that PSO~J334 is most likely not a SMBHB.

\section*{Acknowledgements}
The EVN is a joint facility of independent European, African, Asian, and North American radio astronomy institutes. Scientific results from data presented in this publication are derived from the following EVN project code: RSF08. We thank the Hungarian National Research, Development and Innovation Office (OTKA NN110333) for support. K\'EG was supported by the J\'anos Bolyai Research Scholarship of the Hungarian Academy of Sciences. We thank for the usage of MTA Cloud ({\tt https://cloud.mta.hu}) that significantly helped us achieving the results.

\end{document}